\documentclass[11pt,a4paper]{article}
\pdfoutput=1 

\usepackage{jcappub} 

\usepackage[T1]{fontenc} 

\title{\boldmath Dark matter relic density\\
in Gauss-Bonnet braneworld cosmology
}

\author{Michael T. Meehan}
\author{and Ian B. Whittingham}
\affiliation{College of Science, Technology and Engineering, James Cook University,\\ Townsville 4811, Australia}

\emailAdd{Michael.Meehan@my.jcu.edu.au}
\emailAdd{Ian.Whittingham@jcu.edu.au}

\abstract{The relic density of symmetric and asymmetric dark matter in a Gauss-Bonnet (GB) modified Randall-Sundrum (RS) type II braneworld cosmology is investigated. The existing study of symmetric dark matter in a GB braneworld (Okada and Okada, 2009) found that the expansion rate was reduced compared to that in standard General Relativity (GR), thereby delaying particle freeze-out and resulting in relic abundances which are suppressed by up to $\mathcal{O}(10^{-2})$. This is in direct contrast to the behaviour  observed in RS braneworlds where the expansion rate is enhanced and the final relic abundance boosted. However, this finding that relic abundances are suppressed in a GB braneworld is based upon a highly contrived situation in which the GB era evolves directly into a standard GR era, rather than passing through a RS era as is the general situation. This collapse of the RS era requires equating the mass scale $m_{\alpha}$ of the GB modification and the mass scale $m_{\sigma}$  of the brane tension. However, if the GB contribution is to be considered as the lowest order correction from string theory
to the RS action, we would expect $m_{\alpha} > m_{\sigma}$. We investigate the effect upon the relic abundance of choosing more realistic values for the ratio $\mathcal{R}_{m} \equiv m_{\alpha}/m_{\sigma}$ and find that the relic abundance can be either enhanced or suppressed by more than two orders of magnitude. However, suppression only occurs for a small range of parameter choices and, overwhelmingly, the predominant situation is that of enhancement as we recover the usual Randall-Sundrum type behaviour in the limit $\mathcal{R}_m \gg 1$. We use the latest observational bound $\Omega_{DM}h^2 = 0.1187 \pm 0.0017$ to constrain the various model parameters and briefly discuss the implications for direct/indirect dark matter detection experiments as well as dark matter particle models.}

\begin{document}
\maketitle
\flushbottom
\section{Introduction}
\label{sec:intro}

Precision astrophysical and cosmological measurements have now established that a significant fraction of the matter content in the universe is composed of non-baryonic Dark Matter (DM)~\cite{Hooper}. The data favour cold (non-relativistic) dark matter (CDM) and give the present density as ($68\%$ C.L.)~\cite{Lahav:2014vza}
\begin{equation}
\Omega_{DM} = 0.1187\pm 0.0017\,h^{-2},\label{eq:dm_abun}
\end{equation}
where $\Omega_{DM}$ is the dark matter density as a fraction of the total mass-energy budget and $h = 0.678\pm 0.008$ is defined by the present value of the Hubble constant $H_0 = 100\, h$ km/s/Mpc. The most popular theoretical CDM candidates are WIMPs (Weakly Interacting Massive Particles) with mass $m_{\chi} \sim \mathcal{O}(10 - 1000)$ GeV. One viable WIMP candidate is the neutralino, the lightest supersymmetric particle in supersymmetric extensions of the Standard Model (SM) in which $R$-parity is conserved. 

The origin of the DM can be explained by the thermal relic scenario~\cite{KandT}: at early times, frequent interactions keep the DM particles in equilibrium with the background cosmic bath. As the universe expands and cools, the Boltzmann suppressed interaction rate drops below the expansion rate and the DM particles fall out of equilibrium. At this point - known as particle freeze-out - both annihilation and creation processes cease and the number density redshifts with expansion; the surviving 'relic' particles constitute the dark matter density we observe today. 

Due to the Boltzmann suppression factor in the equilibrium number density, the present dark matter abundance depends sensitively on the timing of freeze-out: the longer a species remains in thermal contact with the background bath, the lower its density at freeze-out. In the standard cosmological model of cold DM with a non-zero cosmological constant (denoted the $\Lambda$CDM model), particle freeze-out occurs during the radiation dominated era when the expansion rate $H \sim T^2/M_{\mathrm{Pl}}$ (where $M_{\mathrm{Pl}} = 1.22\times 10^{19}$ GeV is the Planck mass). In this scenario, a DM candidate with a weak scale interaction cross section, $\sigma \sim G_{\mathrm{F}}^2\,m_\chi^2$, freezes out with an abundance that matches the presently observed value~\eqref{eq:dm_abun} - this is known as the 'WIMP miracle' and strongly motivates thermal WIMP dark matter models. 

Despite the observational success of $\Lambda$CDM, current datasets leave the physics of the universe prior to Big Bang Nucleosynthesis (BBN) ($t \sim 200$ s) relatively unconstrained. If the universe experiences a non-standard expansion law at early times, and in particular during the era of DM decoupling, particle freeze-out may be accelerated (or delayed) and the relic abundance enhanced (or suppressed)~\cite{DBarrow1982501,PhysRevD.81.123522,Pallis:2009ed,Salati2003121,Arbey200846,Gelmini:2013awa,Iminniyaz:2013cla,Meehan:2014zsa} (see also~\cite{Gelmini:2009yh}). 

An interesting class of alternative cosmological models that address this pre-BBN era is provided by the braneworld scenario
in which the observable universe is a 3(+1) dimensional surface (the 'brane') embedded in a five dimensional bulk spacetime. Standard Model particles are confined to the surface of the brane whilst gravity propagates in the higher dimensional bulk~\cite{Langlois:2002bb,Maartens:2010ar}. This class of models is motivated by (super)string theory and M-theory which require additional spacetime dimensions for internal consistency.

In the widely studied Randall-Sundrum type II (RSII) model~\cite{Randall:1999vf}, General Relativity (GR) is recovered on the surface of a 3(+1) Minkowski brane located at the ultraviolet boundary of a five dimensional anti-de Sitter bulk. The warped geometry of the bulk spacetime ensures the fifth dimension is only accessible in the ultraviolet regime and that $\Lambda$CDM is reproduced in the low energy limit. Relic DM abundances in a RSII braneworld model have been investigated for both the case of symmetric DM \cite{PhysRevD.70.083531,PhysRevD.71.063535,PhysRevD.73.063518,PhysRevD.79.115023,
AbouElDahab:2006wb,Meehan:2014zsa}, in which the DM particles are Majorana fermions, that is the particles $\chi $ and antiparticles $\bar{\chi}$ are identical, $\chi = \bar{\chi}$, and the case of asymmetric DM \cite{Meehan:2014zsa} in which the particles and antiparticles are distinct, 
$\chi \ne \bar{\chi}$. In both cases the enhanced early time expansion rate boosts the final relic abundance. 

In this article we consider an extension of the RSII model which incorporates a Gauss-Bonnet (GB) higher order curvature term in the bulk action integral, thus modifying the braneworld dynamics at high energies.\footnote{The inclusion of a GB term affects early universe inflation and modifies both scalar and tensor primordial perturbations and the consistency relation between them~\cite{PhysRevD.70.083525,Tsujikawa2004a,Tsujikawa2004b,Calcagni2013}. Although it produces an enhanced ratio $r$ of the tensor to scalar perturbations~\cite{Neupane:2014}, it is still compatible with the recent Planck~\cite{Ade:2013zuv} and BICEP2~\cite{BICEP2} measurements for the case of single scalar field $m^{2}\phi^{2}$ inflation. For a similar study in the regular Randall-Sundrum model see~\cite{Okada:2014nia}.} The relic density of DM in the Gauss-Bonnet braneworld scenario has been studied by~\cite{PhysRevD.79.103528} for the case of symmetric DM. The GB braneworld effect is treated approximately through the use of a simple multiplicatively modified Hubble expansion which can be interpreted as a multiplicatively modified annihilation cross section in the Boltzmann rate equation and allows the development of an approximate analytic expression for the asymptotic relic abundance. They found that the expansion rate was reduced in the GB model, delaying particle freeze-out and leading to a suppressed relic abundance. This is in direct contrast to the behaviour observed in the RSII braneworld model. This finding, however, is based upon a highly contrived situation in which the Gauss-Bonnet expansion era evolves directly into a standard General Relativity expansion era, rather than passing through a Randall-Sundrum expansion era as is the general case. This collapse of the RS era requires equating the mass scale $m_{\alpha}$ of the GB modification and the mass scale $m_{\sigma}$  of the brane tension.
However, if the GB contribution is to be considered as the lowest order correction from string theory
to the RS action, we would expect $m_{\alpha} > m_{\sigma}$. It is therefore important to investigate the effect upon the relic abundance of choosing more realistic values for the ratio $\mathcal{R}_{m} \equiv m_{\alpha}/m_{\sigma}$ of these two mass scales. 

In the present paper we revisit the calculation of the relic abundance of DM in the GB scenario and study the effects of breaking the assumption $\mathcal{R}_{m}=1$ made by \cite{PhysRevD.79.103528}, replacing it by more realistic values. We also extend the investigation to consider both symmetric and asymmetric DM species and discuss the implications for DM detection experiments and DM particle models.

In the next section we introduce the action integral for the braneworld bulk which includes the Gauss-Bonnet higher curvature term and discuss the modified Friedmann equation in this model. Then, in section~\ref{sec:sdm}, we calculate the DM relic abundance in the Gauss-Bonnet braneworld scenario before deriving constraints on the GB model parameters using the observed relic density. This is repeated for the case of asymmetric DM in section~\ref{sec:adm} and, finally, in section~\ref{sec:con} we summarize our results.

\section{Gauss-Bonnet Braneworlds}

The Randall-Sundrum braneworld model derived from the five dimensional Einstein-Hilbert action can be considered as a low energy effective model of some higher order field theory such as string theory or M-theory. Since our interest in the model lies in the high energy regime where additional quantum corrections in the bulk action may contribute to the braneworld dynamics, we include the leading order correction from heterotic string theory, known as the Gauss-Bonnet term $\mathcal{L}_{GB}$~\cite{Gross198741}, which is given by
\begin{equation}
\mathcal{L}_{GB} = R^2 - 4R_{ab}R^{ab} + R^{abcd}R_{abcd}. 
\end{equation}
Inclusion of higher order curvature terms generally leads to fourth order equations of motion. However,
in five dimensions, the GB combination of invariants constructed from the Riemann tensor $R_{abcd}$ is of particular significance since it is the unique combination that leads to second order gravitational field equations in the bulk metric which are symmetric, divergenceless and ghost free~\cite{Clifton:2011jh}.

Inclusion of the Gauss-Bonnet term modifies the Randall-Sundrum action so that the action integral for the GB braneworld model, taken over the five dimensional bulk spacetime $\mathcal{M}$, is 
\begin{equation}
S_\mathcal{M} = \frac{1}{2\kappa_5^2}\int_\mathcal{M}{d^5x\,\sqrt{-g}\left[R - 2\Lambda_5 + \alpha\mathcal{L}_{GB}\right]},\label{eq:geoact}
\end{equation}
where $g$ is the determinant of the bulk metric $g_{ab}$, $R$ is the five dimensional Ricci scalar and $\Lambda_5(<0)$ is the bulk cosmological constant. We have parameterized the GB contribution through the coupling $\alpha$ which, if this contribution is to be considered as the lowest order correction
from string theory to the Randall-Sundrum action, must 
satisfy~\cite{PhysRevD.70.083525,Tsujikawa2004a} $\alpha |R^{2}| \ll |R|$. Consequently,
$\alpha \ll \ell^{2}$ where $\ell $ is the bulk curvature scale $|R| \propto \ell^{-2}$. 
Introducing the associated energy scale $\mu \equiv \ell^{-1}$ then we require
\begin{equation}
\beta \equiv 4 \alpha \mu ^{2} \ll 1.
\end{equation}
The matter fields, which are localized on the brane surface $\partial\mathcal{M}$, are included via
\begin{equation}
S_m =  - \int_{\partial\mathcal{M}}{d^4x\,\sqrt{-h}\,[\mathcal{L}_{m} + \sigma]},\label{eq:matact}
\end{equation}
where $h$ is the determinant of the induced metric $h_{\mu \nu}$ on the brane surface, 
$\mathcal{L}_m$ is the matter field Lagrangian and $\sigma(>0)$ is the brane tension. 
Varying the total action $S_{tot} = S_\mathcal{M} + S_{m}$ (+ boundary terms) with 
respect to the metric field and solving the resulting field equations yields the modified Friedmann equation for the GB braneworld 
scenario~\cite{0264-9381-19-18-304,PhysRevD.67.024030}
\begin{equation}
\kappa_5^2\left(\rho + \sigma\right) = 2\mu\sqrt{1 + \frac{H^2}{\mu^2}}\left(3 - \beta + 2\beta\frac{H^2}{\mu^2}\right),\label{eq:gb_hub}
\end{equation}
where $\rho$ is the energy density of matter fields on the brane and 
$\beta = 1 - \sqrt{1+ 4\alpha\Lambda_5/3}$. 

The modified Friedmann equation~\eqref{eq:gb_hub} clearly predicts non-standard behaviour for the 
expansion of the universe. However, in the low energy limit, equation~\eqref{eq:gb_hub} reduces to the standard expansion law for a flat universe
\begin{equation}  
H^{2}= \frac{8 \pi}{3 M_{\mathrm{Pl}}^{2}}\rho + \frac{\Lambda_{4}}{3}, \label{eq:st_hub1}
\end{equation}
provided we identify~\cite{Neupane:2001}\footnote{For comparison with~\cite{PhysRevD.70.083531,Meehan:2014zsa}, we note that $\mu$ and $\beta$ are related to the five dimensional Planck mass $M_5$ via
\begin{equation}
M_5^3 = \frac{\mu}{1 + \beta}\frac{M_{\mathrm{Pl}}^2}{8\pi}.\nonumber
\end{equation}}
\begin{equation}
\kappa_4^2 \equiv \frac{8 \pi}{M_{\mathrm{Pl}}^{2}} =  \frac{\mu}{1 + \beta}\kappa_5^2.\label{eq:tune1}
\end{equation}
Additionally, requiring that the four dimensional cosmological constant $\Lambda_{4}$ vanish gives
\begin{equation}
\kappa_5^2 \sigma = 2\mu\left(3 - \beta\right),\label{eq:tune2}
\end{equation}
which is equivalent to the familiar Randall-Sundrum tuning in the limit $\alpha\rightarrow 0$.

As shown in~\cite{PhysRevD.67.103510}, it is possible to solve equation~\eqref{eq:gb_hub} to get an explicit expression for the Hubble factor $H$;
\begin{equation}
H^2 = \frac{\mu^2}{\beta}\left[\left(1 - \beta\right)\cosh{\left(\frac{2\chi}{3}\right)} - 1\right],~\label{eq:gb_hub1}
\end{equation}
where $\chi$ is related to the energy density $\rho$ via
\begin{equation}
\rho + m_\sigma^4 = m_\alpha^4\sinh{\chi},\label{eq:rhomchi}
\end{equation}
and the two mass scales $m_\alpha$ and $m_\sigma$, which correspond to the GB correction and the brane tension respectively, are given by 
\begin{equation}
m_\alpha^4 = \sqrt{\frac{8\mu^2(1 - \beta)^3}{\beta\kappa_5^4}},\quad m_\sigma^4 = \sigma.
\end{equation}
Substituting in the constraints~\eqref{eq:tune1} and~\eqref{eq:tune2}, $m_\alpha$ and $m_\sigma$ can be written in terms of the two remaining free parameters $\mu$ and $\beta$ as
\begin{equation}
m_\alpha^4 = 2\,\frac{\mu^2}{\kappa_4^2}\sqrt{\frac{2(1 - \beta)^3}{\beta(1 + \beta)^2}},\quad m_\sigma^4 = 2\,\frac{\mu^2}{\kappa_4^2}\left(\frac{3-\beta}{1 + \beta}\right).
\end{equation} 
Since the Gauss-Bonnet term is a high energy correction to the regular Randall-Sundrum action, 
we expect $\beta \ll 1$. This motivates us to introduce the quantity
\begin{equation}
\mathcal{R}_{m} \equiv \frac{m_\alpha}{m_\sigma} = \left[\frac{2(1 - \beta)^3}{\beta(3 - \beta)^2}\right]^{1/8},\label{eq:rmdef}
\end{equation}
which measures the ratio of the two mass scales and depends only on $\beta$. The two mass scales
are equal for $\beta = 0.1509$ but, as we expect $\beta \ll 1$, the general situation will be
$\mathcal{R}_m > 1$. 

\begin{figure}[tbp]
\centering 
\includegraphics[scale=0.8,trim=0 0 0 0,clip=true]{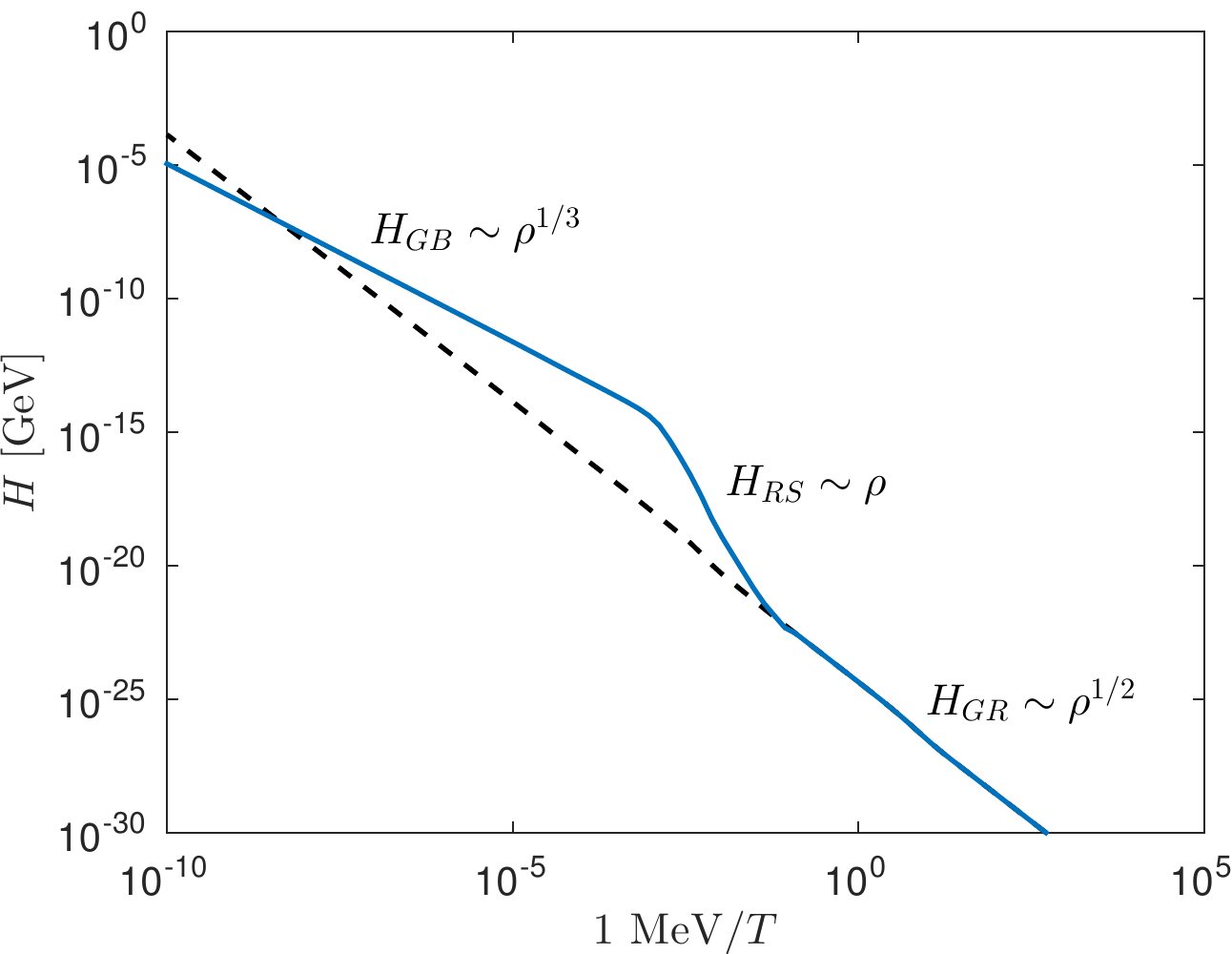}
\caption{\label{fig:GB_hub_betam15_mum44}Modified expansion rate in the Gauss-Bonnet scenario (solid blue curve) for $\mu^2 = 10^{-44}$ GeV$^{2}$ and $\beta = 10^{-15}$. We have assumed that the energy density is radiation dominated for the period shown, taking $\rho = \rho_r = \pi^2\,g_*(T)T^4/30$. The various expansion regimes through which the Hubble parameter evolves are indicated, together with the standard expansion rate (dashed black curve) for reference.}
\end{figure}

Before choosing specific values of $\beta$, we first discuss the evolution of the modified expansion rate in the generalized Gauss-Bonnet scenario. By expanding~\eqref{eq:gb_hub} in the high, intermediate, and low energy limits, we see that the Hubble factor evolves through three distinct expansion regimes, characterized by the mass scales $m_\alpha$ and $m_\sigma$~\cite{PhysRevD.70.083525,PhysRevD.79.103528}:
\begin{enumerate}
\item
The GB regime: $\rho \gg m_\alpha^4$
\begin{equation}
H^2 \simeq \left(\frac{1 + \beta}{4\beta}\mu\kappa_4^2\rho\right)^{2/3},\label{eq:gb_reg}
\end{equation}
\item
The RS regime: $m_\alpha^4\gg \rho \gg m_\sigma^4$
\begin{equation}
H^2 \simeq \frac{\kappa_4^2}{6m_\sigma^4}\rho^2,\label{eq:rs_hub}
\end{equation}
\item
The standard regime: $m_\sigma^4 \gg \rho$
\begin{equation}
H^2 \simeq \frac{\kappa_4^2}{3}\rho.\label{eq:st_hub}
\end{equation}
\end{enumerate}
At early times, during the Gauss-Bonnet regime, the expansion rate of the universe $H \sim \rho^{1/3}$ falls more slowly than the standard expansion law $H\sim \rho^{1/2}$. Later, the universe evolves into a Randall-Sundrum type era with an enhanced expansion $H \sim \rho$, before finally reducing to the standard expansion law in the low energy limit (see figure~\ref{fig:GB_hub_betam15_mum44}).
The duration of the Randall-Sundrum regime is determined by the magnitude of $\mathcal{R}_m \equiv m_\alpha/m_\sigma$: when $\mathcal{R}_m$ is small, the RS era is short and the expansion rate passes quickly from the Gauss-Bonnet era to the standard era; when $\mathcal{R}_m$ is large, the duration of the Randall-Sundrum era is extended. Using the expression for $\mathcal{R}_m$ (equation~\eqref{eq:rmdef}) we see that these two cases correspond to $\beta\lesssim 0.1509$ and $\beta \rightarrow 0$, respectively.

The investigation by~\cite{PhysRevD.79.103528} chose to collapse the Randall-Sundrum era by equating $m_\alpha = m_\sigma$, setting $\beta = 0.1509$. In this case, the early time expansion rate is always slower than (or equal to) the standard expansion rate. The slower expansion rate delays dark matter particle freeze-out and suppresses the relic abundance. This is obviously a contrived scenario considering the Gauss-Bonnet term is a high energy correction to the Randall-Sundrum action and we expect $m_\alpha > m_\sigma$, corresponding to $\beta \ll 1$. In the next section we will show that the unnatural choice of $\beta = 0.1509$ and the conclusions drawn in~\cite{PhysRevD.79.103528} misrepresent the typical behaviour of the relic density in the Gauss-Bonnet braneworld model and that, in fact, the dark matter abundance tends to be enhanced rather than suppressed when realistic values of $\beta$ are used.

It is convenient for the derivation of approximate solutions for the dark matter relic density (see next section) to express the modified expansion rates in the early universe (equations~\eqref{eq:gb_reg} and~\eqref{eq:rs_hub}) in terms of the standard expansion rate $H_{GR}$. 
Since the energy density of the universe during the era of dark matter decoupling is dominated by radiation with $\rho_r = \pi^2 g_*(T) T^4/30$, where $g_*(T)$ is the effective number of relativistic degrees of freedom, equations~\eqref{eq:gb_reg} and~\eqref{eq:rs_hub} can be written as
\begin{align}
H_{GB} &= H_{GR}\left(\frac{x}{x_t^{GB}}\right)^{2/3},\\
H_{RS} &= H_{GR}\left(\frac{x_t^{RS}}{x}\right)^2,
\end{align}
where $x = m_\chi/T$ is a dimensionless variable and $x_t^{GB}$ and $x_{t}^{RS}$ are given by
\begin{align}
\left(x_t^{GB}\right)^4 &\simeq 0.195\,g_*(T_t)m_\chi^4\left(\frac{\beta}{1 + \beta}\right)^2\frac{\kappa_4^2}{\mu^2},\label{eq:xtgb}\\
\left(x_t^{RS}\right)^4 &\simeq 0.082\,g_*(T_t)m_\chi^4\left(\frac{1 + \beta}{3 - \beta}\right)\frac{\kappa_4^2}{\mu^2}.\label{eq:xtrs}
\end{align}

The quantity $x_t^{RS}$ effectively denotes the transition point between the Randall-Sundrum expansion era and the standard expansion era. In order to preserve the successful predictions of BBN, the standard expansion law $H_{GR}$ must be restored prior to $T = 1$ MeV. Thus we require $x_t^{RS}\lesssim 10^3\,m_\chi$, which, using~\eqref{eq:xtrs}, gives the conservative bound 
\begin{equation}
\mu \gtrsim 1\times 10^{-25}\,\mbox{GeV}.\label{eq:murange}
\end{equation}
Furthermore, if we assume that particle freeze-out occurs at $x_f\gtrsim 10$, we can derive an upper limit on the relevant range of $\mu$. Again, using equation~\eqref{eq:xtrs}, we find $\mu \lesssim 5\times 10^{-17}$ GeV and $\mu \lesssim 5\times 10^{-19}$ GeV for $m_\chi = 100$ GeV and $m_\chi = 10$ GeV respectively. For larger values of $\mu$ the standard expansion rate is restored prior to particle freeze-out and particle decoupling is unaffected.

\section{Symmetric Dark Matter}
\label{sec:sdm}
We begin this section by reviewing the relic abundance calculation for a symmetric dark matter species $\chi (= \bar{\chi})$ initially in equilibrium with the background cosmic bath. The dark matter number density $n_\chi$ evolves according to the relativistic Boltzmann equation
\begin{equation}
\frac{dn_\chi}{dt} = -3Hn_\chi - \langle\sigma v\rangle\left(n_\chi^2 - n_\chi^{\mathrm{eq}\,^2}\right),\label{eq:boltsym}
\end{equation}
where $\langle\sigma v\rangle$ is the thermally averaged annihilation cross section and $n_\chi^{\mathrm{eq}}$ is the equilibrium number density. Here, we assume that annihilations are dominated by $s$-wave processes for which the annihilation cross section is a constant, i.e. $\langle\sigma v\rangle = \sigma_0$.\footnote{It is straightforward to extend our analysis to higher partial wave expansions of the annihilation cross section, i.e. $\langle\sigma v\rangle  = \sigma_n x^{-n}$.}

It is convenient to rewrite the Boltzmann equation~\eqref{eq:boltsym} in terms of $x = m_\chi/T$ and the comoving number density $Y=n_\chi/s$, where $s$ is the entropy density given by $s = 2\pi^2 g_*(T) T^4/45$.\footnote{Here $g_*(T)$ actually refers to the number of entropic degrees of freedom $g_{* s}$. Since the number of relativistic and entropic degrees of freedom only differ when a particle crosses a mass threshold, we take $g_{*\rho}=g_{* s} \equiv g_{*}$~\cite{Steigman:2012nb}.} 
We then have
\begin{equation}
\frac{dY}{dx}=-\frac{s\langle\sigma v\rangle}{xH} \zeta(x)\left(Y^2 - Y_{\mathrm{eq}}^2\right),~\label{eq:bolt}
\end{equation}
where $Y_{\mathrm{eq}} \simeq 0.145(g_\chi/g_*)\,x^{3/2}e^{-x}$, $g_\chi = 2$ is the number of internal degrees of freedom of the dark matter species $\chi$ and
\begin{equation} 
\zeta (x) = 1 - \frac{1}{3}\frac{d\log{g_*}}{d\log{x}}~\label{eq:df} 
\end{equation}
is a temperature dependent factor related to the change in the number of degrees of freedom.
The present dark matter density, $\Omega_{DM}h^2$, is obtained from the asymptotic 
solution ($x\rightarrow\infty$) of equation~\eqref{eq:bolt}
\begin{equation}
\Omega_{DM}h^2 = 2.75\times 10^8\,m_\chi Y_\infty,\label{eq:omgdm}
\end{equation}
where $Y_\infty=Y(x \rightarrow \infty)$ is the present comoving density. 

In general, the Boltzmann equation cannot be solved analytically and equation~\eqref{eq:bolt} must be integrated numerically. However, an approximate solution can be found by exploiting the exponential decay of $Y_{\mathrm{eq}}$: as outlined in~\cite{PhysRevD.33.1585,KandT}, the creation term ($\propto Y_{\mathrm{eq}}^2$) in
equation~\eqref{eq:bolt} can be neglected following particle decoupling (i.e. for $x > x_f$) and the resulting expression can be integrated directly once the expansion rate and annihilation cross section have been specified. Taking $H = H_{GR}$ and $\langle\sigma v \rangle =$ constant, we get the well-known approximate solution for the asymptotic comoving density in the standard cosmological scenario\footnote{The annihilation cross sections $\langle\sigma v\rangle$ in equations~\eqref{eq:yinfst},~\eqref{eq:ysym} and~\eqref{eq:yinfrs} have units of GeV$^{-2}$ to match the units of $\lambda$.}
\begin{equation}
Y_\infty^{GR} \simeq \frac{x_f^{GR}}{\lambda_{GR}\langle\sigma v\rangle},\label{eq:yinfst}
\end{equation}
where $\lambda_{GR} \simeq 0.264 \sqrt{g_*}M_{\mathrm{Pl}}\,m_\chi$ and $x_f^{GR}$ is the freeze-out point in the standard scenario that can be estimated using~\cite{PhysRevD.33.1585,KandT}
\begin{equation}
x_f^{GR} \simeq \log{\left[\left(2 + c\right)\lambda_{GR}\langle\sigma v\rangle ac\right]} - \frac{1}{2}\log{\left\{\log{\left[\left(2 + c\right)\lambda_{GR}\langle\sigma v\rangle ac\right]}\right\}},
\end{equation}
with $a \simeq 0.145\,g_\chi/g_*$ and $c \approx 0.6$ a numerical constant (see~\cite{KandT} for more details).\footnote{In deriving equation~\eqref{eq:yinfst} the number of relativistic degrees of freedom has been fixed at $g_*(T) = g_*(T_f)$, but, note that the full temperature dependence is restored in the numerical integration.}

In the Gauss-Bonnet braneworld scenario, the universe first passes through a Gauss-Bonnet and then a Randall-Sundrum type expansion era before relaxing to the standard expansion law (see previous section). We therefore need to find equivalent expressions to~\eqref{eq:yinfst} for when dark matter decoupling occurs during each of these non-standard regimes. 

Taking $H = H_{GB}$ (equation~\eqref{eq:gb_hub}), we find that if decoupling occurs during a Gauss-Bonnet type expansion regime~\cite{PhysRevD.79.103528},
\begin{equation}
Y_{\infty}^{GB} \simeq \frac{5}{3}\frac{(x_f^{GB})^{5/3}}{\lambda_{GB}\langle\sigma v\rangle},\label{eq:ysym}
\end{equation}
where
\begin{align}
\lambda_{GB} &= \lambda_{GR}\left(x_t^{GB}\right)^{2/3}\nonumber\\
&\simeq \left[\left(\frac{\beta}{1 + \beta}\right)g_*^2\frac{m_\chi^5}{\mu\kappa_4^2}\right]^{1/3},\label{eq:lamgb}
\end{align}
and the freeze-out point is
\begin{equation}
x_f^{GB} \simeq \log{\left[\left(2 + c\right)\lambda_{GB}\langle\sigma v\rangle ac\right]}  - \frac{7}{6}\log{\left\{\log{\left[\left(2 + c\right)\lambda_{GB}\langle\sigma v\rangle ac\right]}\right\}}.
\end{equation}
Similarly, if decoupling occurs during the Randall-Sundrum era~\cite{PhysRevD.70.083531}
\begin{equation}
Y_\infty^{RS} \simeq \frac{0.54\,x_t^{RS}}{\lambda_{GR}\langle\sigma v\rangle},\label{eq:yinfrs}
\end{equation}
which we note is independent of the freeze-out point $x_f$ (provided $x_t^{RS} \gg x_f$). 

Comparing equations~\eqref{eq:ysym} and~\eqref{eq:yinfrs} with~\eqref{eq:yinfst}, we see that the asymptotic comoving density can be either suppressed or enhanced depending on the relative magnitude of $\mu$ and $\beta$ and the timing of particle decoupling. More specifically, if decoupling occurs during the Gauss-Bonnet era, the comoving density may be either enhanced or suppressed, otherwise, if decoupling occurs during the Randall-Sundrum era, the comoving density is always enhanced.

To determine which parameter combinations lead to suppression, and which lead to enhancement (with respect to the standard cosmology result) we can equate equations~\eqref{eq:ysym} and~\eqref{eq:yinfrs} with~\eqref{eq:yinfst}. Rearranging for $\mu^2$, we find that the relic abundance is enhanced for the interval\footnote{To derive~\eqref{eq:enhint} and~\eqref{eq:supint} we have assumed that the freeze-out point is roughly constant. In doing so we have neglected a logarithmic dependence on the annihilation cross section $\langle\sigma v\rangle$.}
\begin{equation}
5\times 10^{-43}\,m_\chi^4\left(\frac{\beta}{1 + \beta}\right)^2\lesssim \mu^2 \lesssim 1\times 10^{-41}\,m_\chi^4,\label{eq:enhint}
\end{equation}
and suppressed for
\begin{equation}
\mu^2 \lesssim 5\times 10^{-43}\,m_\chi^4\left(\frac{\beta}{1 + \beta}\right)^2.\label{eq:supint}
\end{equation}
For $\mu^2 \gtrsim 10^{-41} m_\chi^4$, the standard expansion rate is restored prior to particle decoupling and the predicted value of $\Omega_{DM}h^2$ reduces to the canonical result.

\begin{figure}[tbp]
\centering 
\includegraphics[scale=0.57,trim=0 0 0 0,clip=true]{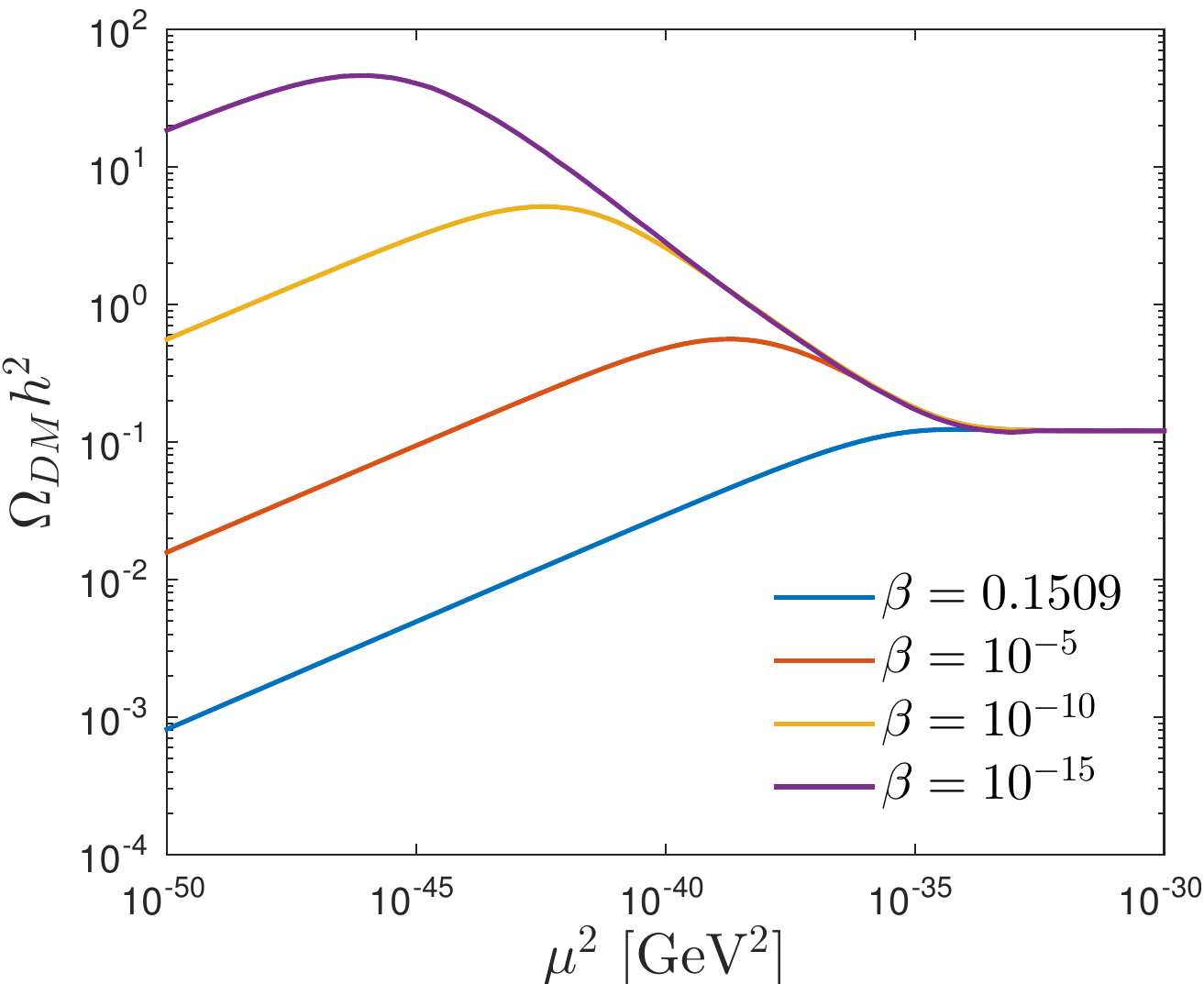}
\hfill
\includegraphics[scale=0.57,trim=0 0 0 0,clip=true]{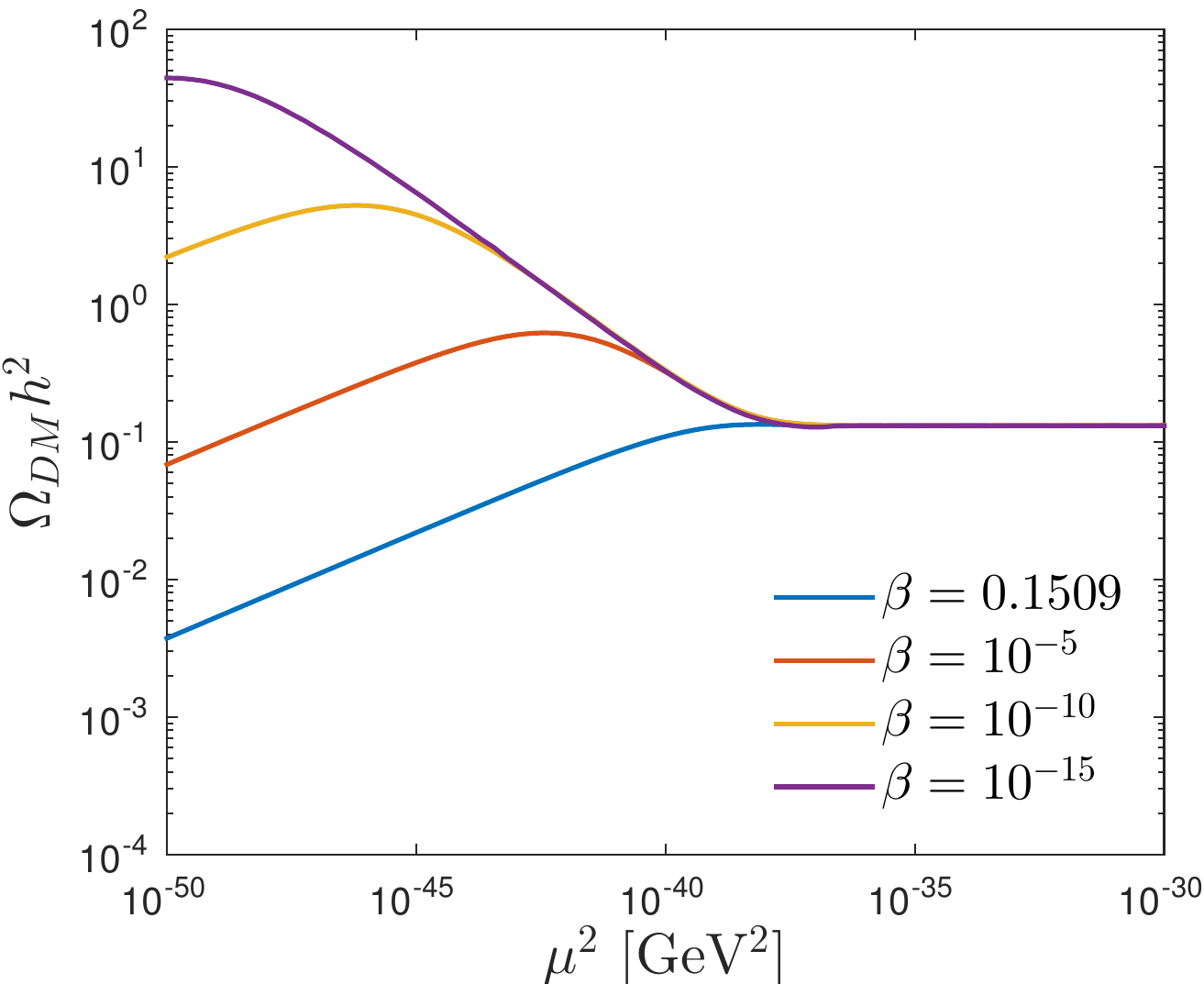}
\caption{\label{fig:musq_omega_varbeta} Relic abundance $\Omega_{DM}h^2$ for a symmetric WIMP with 
$\langle\sigma v\rangle = 2\times 10^{-26}$ cm$^3$s$^{-1}$ as a function of $\mu^2$ for $\beta = 0.1509$ (blue curve), $\beta = 10^{-5}$ (red curve), $\beta = 10^{-10}$ (yellow curve) and $\beta = 10^{-15}$ (purple curve). The left and right panels correspond to WIMP masses $m_\chi = 100$ GeV and 10 GeV respectively.}
\end{figure}

In figure~\ref{fig:musq_omega_varbeta} we plot the predicted relic abundance $\Omega_{DM}h^2$ in the general Gauss-Bonnet scenario as a function of $\mu^2$ for varying $\beta$. 
Immediately we see that $\Omega_{DM}h^2$ (much like the expansion rate $H$) can be split up into three distinct regions: for small $\mu^2$ (and large $\beta$), the relic density increases with increasing $\mu^2$ (and decreasing $\beta$), reaching a maximum that is approximately given by\footnote{The parameter dependence of the maximum can be derived by equating~\eqref{eq:ysym} with~\eqref{eq:yinfrs}. Note, however, that the numerical constants are only approximate because we have not taken into account the variation in $x_f$.}
\begin{equation}
\Omega^{\mathrm{max}}_{DM}h^2 \sim \frac{9\times 10^{-11}}{\beta^{1/5}\langle\sigma v\rangle};\qquad \mu^2_{\mathrm{max}} \sim 3\times 10^{-43}\,(m_\chi \beta^{1/5})^{4}\,\,\mbox{GeV}^2.\label{eq:max}
\end{equation}
In this region, decoupling occurs during the Gauss-Bonnet expansion era and the relic density can be estimated using~\eqref{eq:ysym}. Next, for $\mu^2 \gtrsim \mu^2_{\mathrm{max}}$, the relic density decreases with increasing $\mu^2$ and is relatively independent of $\beta$. Here, decoupling occurs during the Randall-Sundrum era and each curve approaches the Randall-Sundrum result~\cite{PhysRevD.70.083531}. Finally, when $\mu^2 \gtrsim 10^{-41} m_\chi^4$, each curve reduces to the standard cosmology result. Hence, for the purpose of estimating the relic density, three approximate regimes can be
identified:
\begin{align}
\mu^2 &\lesssim 3\times 10^{-43}m_\chi^4\beta^{4/5} &:\quad \mbox{GB regime}\label{eq:gb_freeze}\\
3\times 10^{-43}m_\chi^4\beta^{4/5} \lesssim \mu^2 &\lesssim 10^{-41}m_\chi^4 &: \quad \,\mbox{RS regime}\label{eq:rs_freeze}\\
\mu^2 &\gtrsim 10^{-41}m_\chi^4 &: \quad \mbox{GR regime}\label{eq:st_freeze}
\end{align}
within which equations~\eqref{eq:ysym},~\eqref{eq:yinfrs} and~\eqref{eq:yinfst} for $Y_{\infty}$ would be appropriately used.

As expected, figure~\ref{fig:musq_omega_varbeta} shows that the dark matter relic abundance may be either enhanced or suppressed by up to two or more orders of magnitude, depending on the values of $\mu^2$ and $\beta$. We must stress, however, that as the value of $\beta$ is reduced, the predicted relic density tends towards the Randall-Sundrum result, and is therefore enhanced. Also, since $\mu^2 \gtrsim 10^{-50}$ GeV$^2$ is bounded from below by BBN constraints, suppression is only possible if $\beta\gtrsim 1.4\times 10^{-4}/m_\chi^2$, corresponding to the condition $\mathcal{R}_m \lesssim 3.3\,m_\chi^{1/4}$. Furthermore, it is only for the particular case considered in~\cite{PhysRevD.79.103528}, that is $\beta = 0.1509$ ($\mathcal{R}_m = 1$) (blue curve), that $\Omega_{DM}h^2$ is exclusively suppressed. For more reasonable values of $\beta$ (and $\mathcal{R}_m$) the relic density is typically enhanced.

\begin{figure}[tbp]
\centering 
\includegraphics[scale=0.48,trim=0 0 0 0,clip=true]{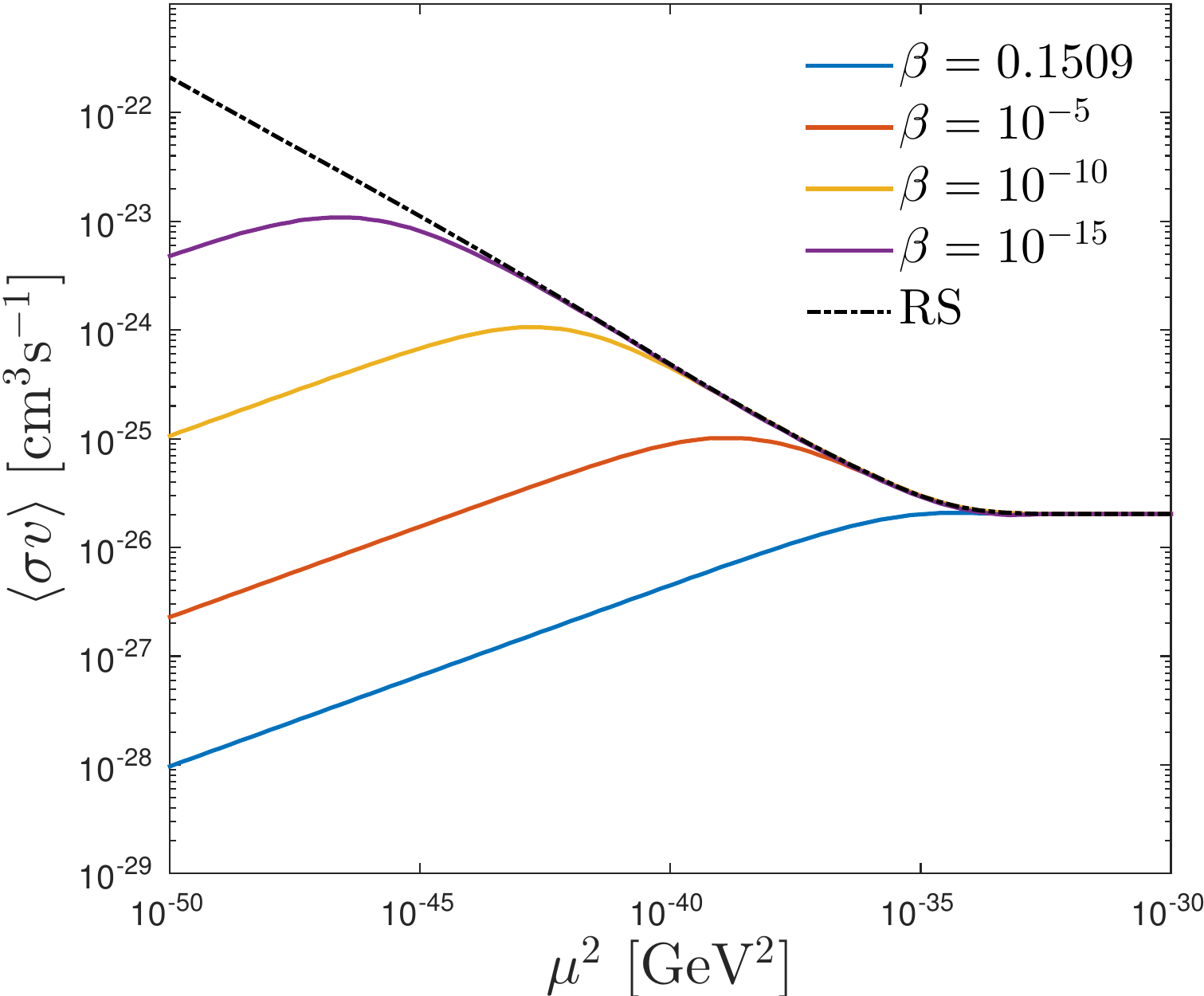}
\hfill
\includegraphics[scale=0.48,trim=0 0 0 0,clip=true]{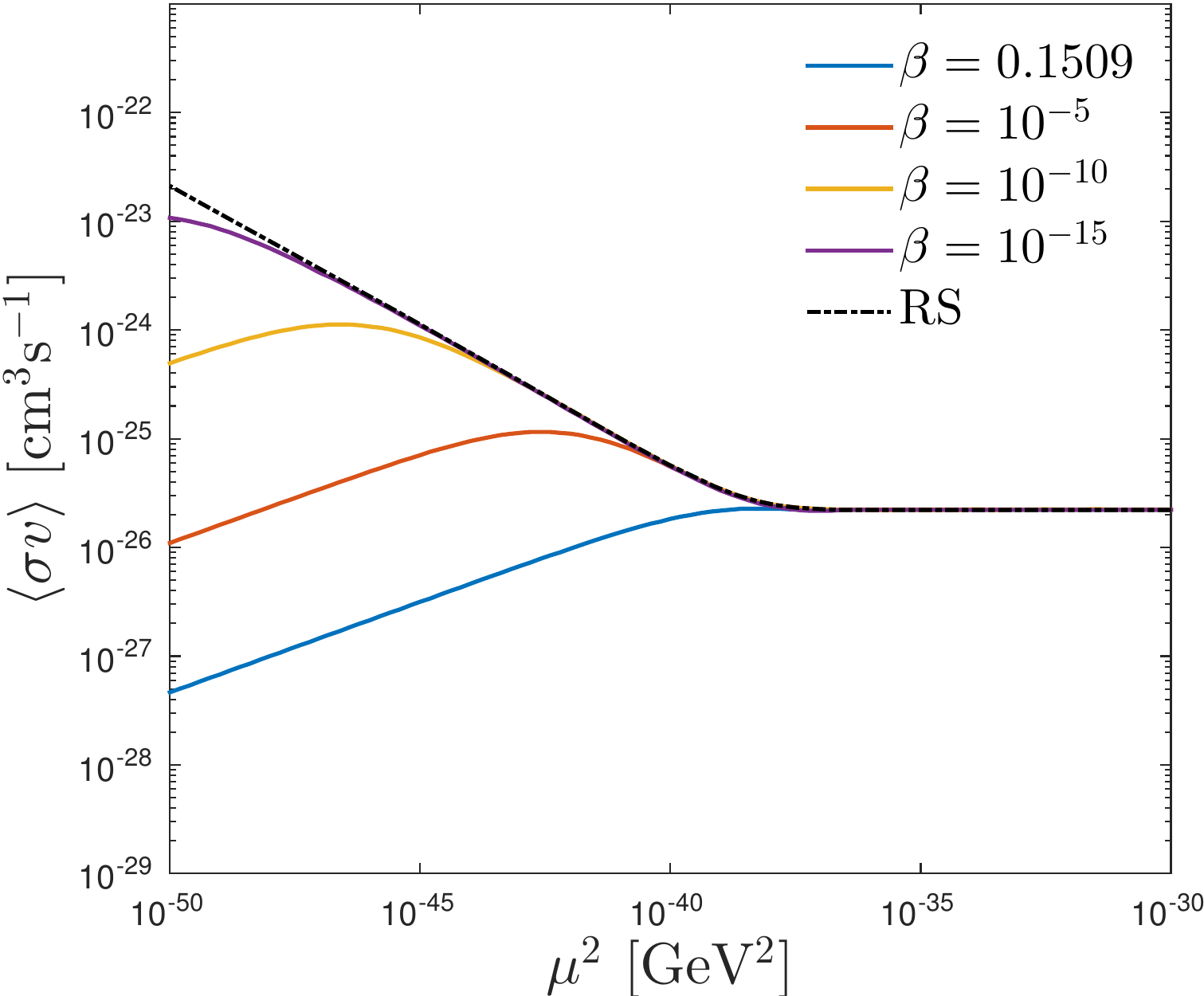}
\caption{\label{fig:musq_sigma_cont_varbeta} Required annihilation cross section $\langle\sigma v\rangle$ for a symmetric WIMP as a function of $\mu^2$ for $\beta = 0.1509$ (blue curve), $\beta = 10^{-5}$ (red curve), $\beta = 10^{-10}$ (yellow curve) and $\beta = 10^{-15}$ (purple curve). Also shown is the corresponding result for a pure Randall-Sundrum scenario (dot-dashed black curve). The left and right panels correspond to WIMP masses $m_\chi = 100$ GeV and 10 GeV respectively.}
\end{figure}

We can invert these results to find the annihilation cross section required to produce the observed relic density $\Omega_{DM}h^2 = 0.1187$. In figure~\ref{fig:musq_sigma_cont_varbeta} we plot this cross section as a function of $\mu^2$ for varying $\beta$. The cross section, which is inversely proportional to $\Omega_{DM}h^2$, exhibits similar behaviour to the relic density curves presented in figure~\ref{fig:musq_omega_varbeta} in that the three regimes - Gauss-Bonnet, Randall-Sundrum and standard - are immediately apparent.

The required cross section in each regime can be estimated by rearranging the approximate expressions~\eqref{eq:ysym},~\eqref{eq:yinfrs} and~\eqref{eq:yinfst} and substituting in the observed relic density $\Omega_{DM}h^2$. Thus, if decoupling occurs deep in the Gauss-Bonnet era, the required annihilation cross section is given by
\begin{equation}
\langle\sigma v\rangle \simeq 2.0\times 10^{-22}\left(\frac{1 + \beta}{\beta}\,\frac{\mu}{m_\chi^2}\right)^{1/3}\frac{\left(x_f^{GB}\right)^{5/3}}{\Omega_{DM}h^2}\quad\mbox{cm}^3\mbox{s}^{-1}.\label{eq:sigmagb}
\end{equation}
Similarly, if decoupling occurs during the Randall-Sundrum era,
\begin{equation}
\langle\sigma v\rangle \simeq 9.4\times 10^{-38}\left[\left(\frac{1 + \beta}{3 - \beta}\right)\frac{1}{\mu^2}\right]^{1/4}\frac{m_\chi}{\Omega_{DM}h^2}\quad\mbox{cm}^3\mbox{s}^{-1},\label{eq:sigmars}
\end{equation}
which is relatively independent of $\beta$. For $\mu^2\gtrsim 10^{-41}m_\chi^4$, the transition point $x_t^{RS}$ precedes the freeze-out point and we recover the canonical result $\langle\sigma v\rangle \simeq \langle\sigma v\rangle^{GR} \simeq 2.03\times 10^{-26}$ cm$^3$s$^{-1}$ and $\langle\sigma v\rangle \simeq \langle\sigma v\rangle^{GR} \simeq 2.21\times 10^{-26}$ cm$^3$s$^{-1}$ for $m_\chi = 100$ GeV and $m_\chi =10$ GeV respectively.\footnote{Note that the approximate expressions~\eqref{eq:sigmagb} and~\eqref{eq:sigmars} are more accurate than the corresponding expressions involving the relic density since there is much less variation in the freeze-out point $x_f$ once $\Omega_{DM}h^2$ has been specified.}

The results in figure~\ref{fig:musq_sigma_cont_varbeta} should be compared with the latest constraints derived from the Fermi-LAT gamma ray data~\cite{Fermi}. For example, the bounds for the $\chi\bar{\chi}\rightarrow b\bar{b}$ and $\chi\bar{\chi}\rightarrow \mu^+\mu^-$ annihilation channels for a dark matter particle with mass $m_\chi = 100$ GeV are $\langle\sigma v\rangle_{\mathrm{Fermi}} = 1.31\times 10^{-25}$ cm$^3$s$^{-1}$ and $\langle\sigma v\rangle_{\mathrm{Fermi}} = 1.38\times 10^{-24}$ cm$^3$s$^{-1}$ respectively. For the $m_\chi = 10$ GeV case the bounds are more stringent with $\langle\sigma v\rangle_{\mathrm{Fermi}} = 2.90\times 10^{-26}$ cm$^3$s$^{-1}$ and $\langle\sigma v\rangle_{\mathrm{Fermi}} = 2.01\times 10^{-25}$ cm$^3$s$^{-1}$ for the respective channels. The Fermi-LAT constraints therefore exclude a portion of the Gauss-Bonnet parameter space.  For the small values of $\beta$ that correspond to realistic values $\mathcal{R}_m > 1$, larger values of $\mu^2$ are favoured. We must keep in mind however, that these constraints only apply if the dark matter particle annihilates primarily through one of the channels mentioned.

\section{Asymmetric Dark Matter}
\label{sec:adm}

Asymmetric dark matter models treat the dark matter particle $\chi$ and antiparticle $\bar{\chi}$ as distinct and with unequal number densities, similar to the asymmetry that exists in the baryonic sector. In fact, these models typically assume~\cite{Kumar,Graesser:2011wi} either a primordial asymmetry in one sector that is transferred to the other sector, or that both asymmetries are generated by the same physical process such as the decay of a heavy particle. Connecting the two asymmetries also explains the proximity of the dark and baryonic densities, $\Omega_{DM}/\Omega_b \sim 5$, suggesting the dark matter mass is in the range $m_\chi \sim 5 - 15$ GeV~\cite{PhysRevD.79.115016}.

When the particle $\chi$ and antiparticle $\bar{\chi}$ are distinct, the Boltzmann equation~\eqref{eq:boltsym} is generalized to the coupled system
\begin{subequations}
\begin{eqnarray}
\frac{dn_\chi}{dt} &=- 3Hn_\chi -\langle\sigma v\rangle\left(n_\chi n_{\bar{\chi}} - n_\chi^{\mathrm{eq}}n_{\bar{\chi}}^{\mathrm{eq}}\right),\label{eq:nchi} \\
\frac{dn_{\bar{\chi}}}{dt} &= - 3Hn_{\bar{\chi}} -\langle\sigma v\rangle\left(n_\chi n_{\bar{\chi}} - n_\chi^{\mathrm{eq}}n_{\bar{\chi}}^{\mathrm{eq}}\right)\label{eq:nchibar},
\end{eqnarray}
\label{eq:nchiasym}
\end{subequations}
where $n_\chi^{\mathrm{eq}}$ and $n_{\bar{\chi}}^{\mathrm{eq}}$ are the equilibrium number densities of the $\chi$ and $\bar{\chi}$ components respectively.
We assume that self annihilations are forbidden, and that only interactions of the type $\chi\bar{\chi}\rightarrow X\bar{X}$ (where the $X$'s are Standard Model particles) can change the dark matter particle number. We can then write
\begin{equation}
Y_\chi - Y_{\bar{\chi}} = C\label{eq:Cdef},
\end{equation}
where $C$ is a strictly positive constant that characterizes the asymmetry between the particles and antiparticles. Here, we are not concerned with the mechanism that generates the asymmetry, only that one has been created well before particle freeze-out.

Rewriting the Boltzmann equations in terms of the comoving density $Y$, and using equation~\eqref{eq:Cdef}, the system~\eqref{eq:nchiasym} becomes
\begin{align}
\frac{dY_{\chi}}{dx}&=-\frac{s\langle\sigma v\rangle}{xH}\zeta(x)\left(Y_\chi^2 - CY_{\chi} - P\right),\nonumber\\
\frac{dY_{\bar{\chi}}}{dx}&=-\frac{s\langle\sigma v\rangle}{xH}\zeta(x)\left(Y_{\bar{\chi}}^2 + CY_{\bar{\chi}} - P\right),\label{eq:dYasym}
\end{align}
where, since the dark matter particles and antiparticles are non-relativistic at decoupling,
\begin{equation}
P \equiv Y_\chi^{\mathrm{eq}}Y_{\bar{\chi}}^{\mathrm{eq}} = \left(\frac{0.145\,g_\chi}{g_{*}}\right)^2x^3e^{-2x}.\label{eq:Pdef}
\end{equation}
Solving the system~\eqref{eq:dYasym} in the asymptotic limit, the total dark matter density, $\Omega_{DM}h^2$, is the sum of the $\chi$ and $\bar{\chi}$ components,
\begin{equation}
\Omega_{DM}h^2 = 2.75\times 10^8\,m_\chi\left(Y_\chi^\infty + Y_{\bar{\chi}}^\infty\right).\label{eq:dmtot}
\end{equation}
Following similar arguments to those for the symmetric case, we can find an approximate 
solution to the system~\eqref{eq:dYasym} for the asymptotic density of the $\bar{\chi}$ 
component (see~\cite{Iminniyaz:2011yp} for details)
\begin{equation}
Y_{\bar{\chi}}^\infty \simeq \frac{C}{\exp{\left(C/Y^\infty_{(sym)}\right)} - 1},\label{eq:ychibarapp}
\end{equation}
where we use $Y^\infty_{(sym)}$ to denote the corresponding asymptotic solution for symmetric dark matter. As we saw in the previous section, $Y^\infty_{(sym)}$ depends on the timing of freeze-out and
is given by, respectively, equations~\eqref{eq:yinfst},~\eqref{eq:ysym} and~\eqref{eq:yinfrs} for the 
three regimes~\eqref{eq:gb_freeze}-\eqref{eq:st_freeze}. From equations~\eqref{eq:ychibarapp} 
and~\eqref{eq:Cdef}, we readily obtain 
\begin{equation}
Y_{\chi}^\infty \simeq \frac{C}{1 - \exp{\left(-C/Y^\infty_{(sym)}\right)}}.\label{eq:ychiapp}
\end{equation}

As discussed in~\cite{Iminniyaz:2011yp,Meehan:2014zsa}, the contribution from the minority and majority components to the total dark matter density depends sensitively on the ratio $C/Y^{\infty}_{(sym)}$.  When $C/Y_{(sym)}^\infty\gg 1$, the density of the $\bar{\chi}$ component is exponentially suppressed, $Y_{\bar{\chi}}^\infty \simeq C\,\exp{\left(-C/Y^\infty_{(sym)}\right)}$, and the density of the $\chi$ component approaches the asymmetry $C$, $Y_\chi\simeq C + C\,\exp{\left(-C/Y^{\infty}_{(sym)}\right)}$. Conversely, when $C/Y_{(sym)}^\infty \ll 1$, the factor $C$ drops out of the expressions~\eqref{eq:ychibarapp} and \eqref{eq:ychiapp} and each component behaves like symmetric dark matter, i.e. $Y_\chi^\infty \simeq Y_{\bar{\chi}}^\infty \simeq Y^\infty_{(sym)}$. We designate each of these regimes as being \textit{strongly} and \textit{weakly} asymmetric respectively, with the relic density in each case behaving like
\begin{equation}
\Omega_{DM}h^2 \simeq 
\begin{cases}
\,2\times 2.75\times 10^8\,m_\chi\,Y_{(sym)}^\infty, & \quad C/Y_{(sym)}^\infty\ll 1,\\
\,2.75\times 10^8\,m_\chi\,C, & \quad C/Y_{(sym)}^\infty\gg 1.\label{eq:omglim}
\end{cases}
\end{equation}

To determine which parameter values correspond to each regime, we use the results derived in the previous section for symmetric dark matter. There we saw that the relic density was enhanced for the interval~\eqref{eq:enhint},
\begin{equation}
5\times 10^{-43}\,m_\chi^4\left(\frac{\beta}{1 + \beta}\right)^2\lesssim \mu^2 \lesssim 1\times 10^{-41}\,m_\chi^4,
\end{equation}
and suppressed for the interval~\eqref{eq:supint}
\begin{equation}
\mu^2 \lesssim 5\times 10^{-43}\,m_\chi^4\left(\frac{\beta}{1 + \beta}\right)^2.
\end{equation}
Therefore, for a fixed value of the asymmetry $C$, these two cases would drive the dark matter species towards the weakly or strongly asymmetric regimes respectively. 

Again, we can invert the expressions for the asymptotic comoving densities~\eqref{eq:ychibarapp} and~\eqref{eq:ychiapp} and, using~\eqref{eq:dmtot}, find the annihilation cross section required to produce the observed relic density. Then, depending on the timing of freeze-out (see equations~\eqref{eq:gb_freeze}-\eqref{eq:st_freeze}), the cross section and asymmetry are related via
\begin{equation}
   \langle\sigma v\rangle \simeq \frac{a}{C}\coth^{-1}\left(\frac{\omega}{C}\right) \times
     \begin{cases}
       10\,\left(x_f^{GB}\right)^{5/3}/(3\lambda_{GB}) &;  \quad (\mbox{GB}\,\mbox{regime})\\
       1.1\,x_t^{RS}/\lambda_{GR}       &;  \quad (\mbox{RS}\,\mbox{regime})\\
			 2\,\left(x_f^{GB}\right)/\lambda_{GR}         &;  \quad (\mbox{GR}\,\mbox{regime})\label{eq:sigma_asym}\\
     \end{cases}
\end{equation}
where $\omega = \Omega_{DM}h^2/(2.75\times 10^8 m_\chi)$ and $a=1.167\times 10^{-17} \mbox{cm}^3\mbox{s}^{-1}$.

\begin{figure}[tbp]
\centering
\includegraphics[scale = 0.85, trim = 0 0 0 0, clip = true]{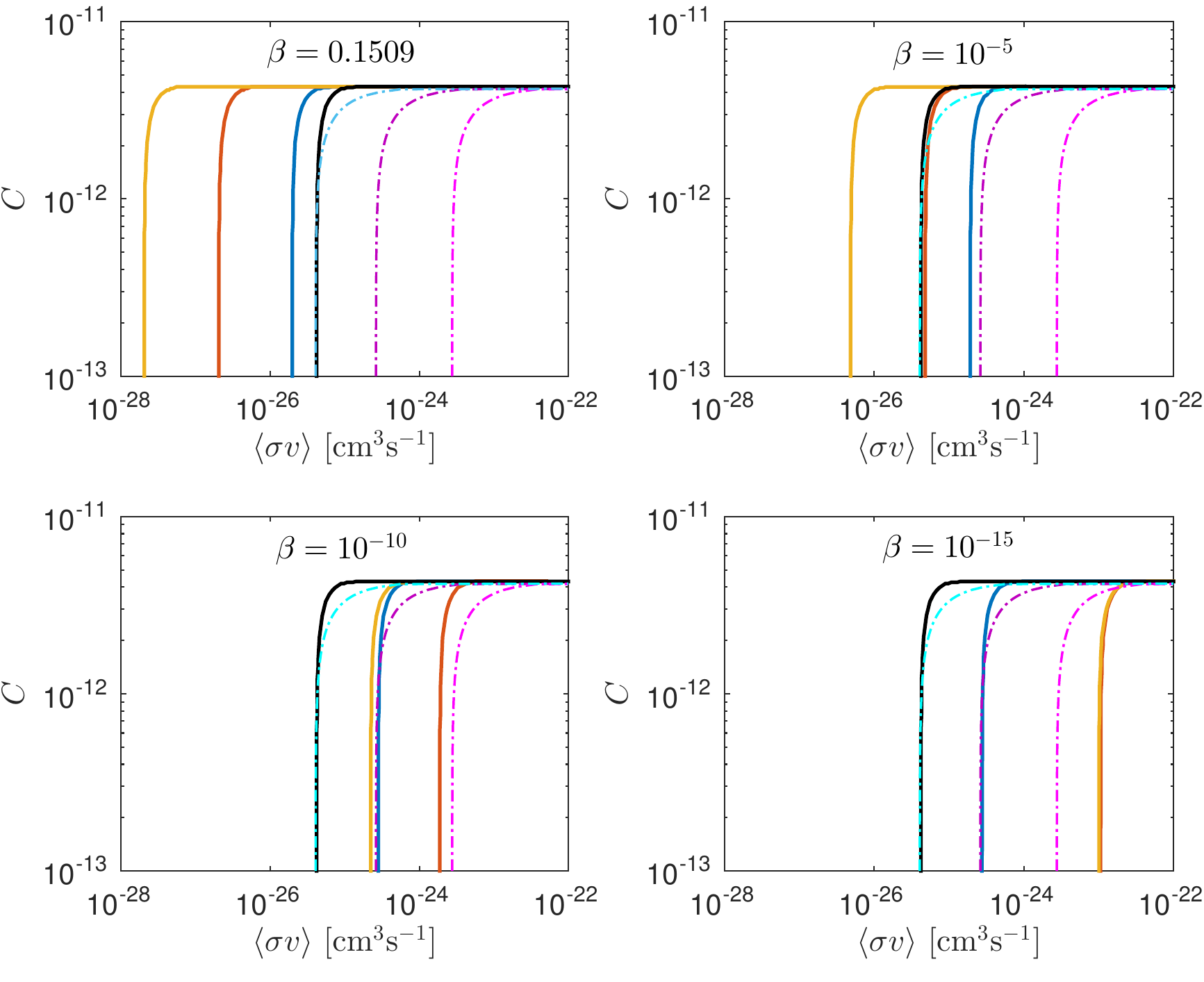}
\caption{\label{fig:sig_C_cont_varbeta100} Iso-abundance contours in the $(\langle\sigma v\rangle, C)$ plane corresponding to the observed dark matter abundance $\Omega_{DM}h^2 = 0.1187$ for a $100$ GeV WIMP. The  contours shown are for $\mu^2 = 10^{-38}$ GeV$^2$ (solid blue curve), 
$\mu^2 = 10^{-44}$ GeV$^2$ (solid red curve) and $\mu^2 = 10^{-50}$ GeV$^2$ (solid yellow curve). Also shown is the standard cosmology result (solid black curve). The panels correspond to $\beta = 0.1509$ (top left), $\beta = 10^{-5}$ (top right), $\beta = 10^{-10}$ (bottom left) and $\beta = 10^{-15}$ (bottom right). Note that, for $\beta = 10^{-15}$, the contours for $\mu^2 = 10^{-44}$ GeV$^2$ and $\mu^2 = 10^{-50}$ GeV$^2$ (almost) coincide. In each panel we have superimposed the constraints derived from the Fermi-LAT gamma ray data~\cite{Fermi} with the regions below the dark purple and magenta (dot-dashed) curves excluded for the $\mu^+\mu^-$ and $b\bar{b}$ annihilation channels respectively. We have also indicated the region (below the dot-dashed blue curve) for which the asymmetric detection signal in the Gauss-Bonnet scenario exceeds the symmetric signal in the standard scenario.}
\end{figure}

\begin{figure}[tbp]
\centering
\includegraphics[scale = 0.85, trim = 0 0 0 0, clip = true]{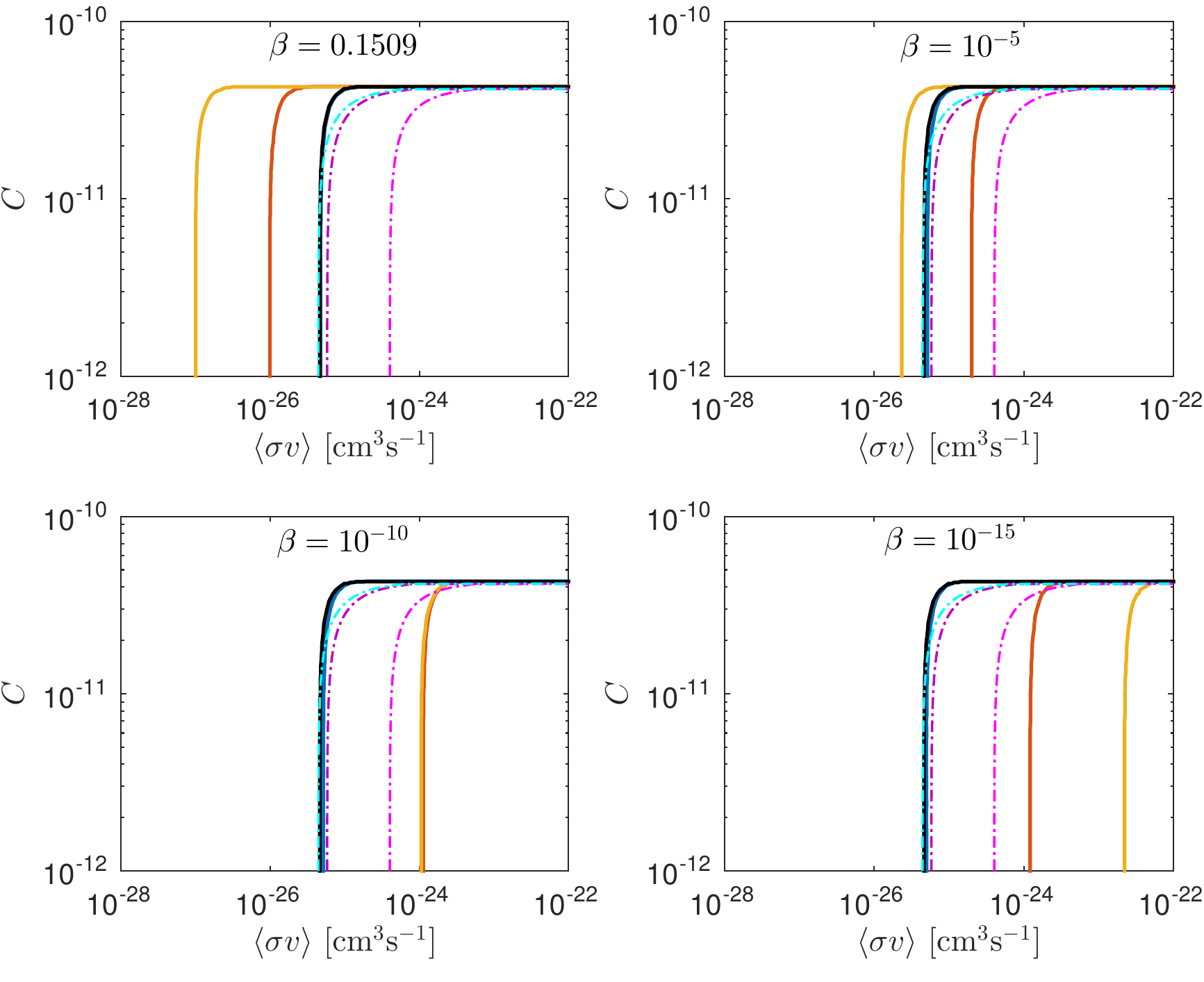}
\caption{\label{fig:sig_C_cont_varbeta10} Same as figure~\ref{fig:sig_C_cont_varbeta100} but for $m_\chi = 10$ GeV. In each panel the contour for $\mu^2 = 10^{-38}$ GeV$^2$ (almost) overlaps the standard cosmology result.}
\end{figure}

The numerical results for the required annihilation cross section are plotted in figures~\ref{fig:sig_C_cont_varbeta100} and~\ref{fig:sig_C_cont_varbeta10} (solid curves) for $m_\chi = 100$ GeV and $m_\chi = 10$ GeV respectively. The different curves within each panel correspond to different values of $\mu^2$ and we have reduced the magnitude of $\beta$ in the successive panels. In each figure we plot the standard cosmology result (black) for reference.

Initially the curves are vertical and the relic density is determined solely by the annihilation cross section. In this region the ratio $C/Y^\infty_{(sym)}$ is small and each component behaves like symmetric dark matter. As both the annihilation cross section and the asymmetry increase we transition into a regime which is strongly asymmetric where the curves are horizontal. Here the density of the minority component is exponentially suppressed and the relic abundance is fixed by the asymmetry $C$. This general behaviour is exhibited regardless of the values of $\mu^2$ or $\beta$, however, the magnitude of the annihilation cross section which separates the weakly and strongly asymmetric regions depends significantly on the combination of $\mu^2$ and $\beta$ (see~\eqref{eq:sigma_asym}). 

Since the vertical section of each curve corresponds to the weakly asymmetric regime, the position of the vertical asymptotes can be deduced simply from figure~\ref{fig:musq_sigma_cont_varbeta} (with allowance for the additional factor of $\sim 2$ due to the $\chi$ and $\bar{\chi}$ contributions). When the annihilation cross section is enhanced in figure~\ref{fig:musq_sigma_cont_varbeta}, the curves in figures~\ref{fig:sig_C_cont_varbeta100} and~\ref{fig:sig_C_cont_varbeta10} will be shifted to the right of the standard cosmology result. Similarly, when the symmetric cross section is suppressed, the asymmetric curves will be shifted towards the left. Thus the symmetric cross section determines the vertical asymptote of the required asymmetric cross section.

Consequently, just like the symmetric case, the required annihilation cross section is reduced for all values of $\mu^2$ when $\beta = 0.1509$ (panel 1), getting smaller with decreasing $\mu^2$. Then, as the magnitude of $\beta$ is decreased (in successive panels), the curves are shifted towards larger cross sections. There is a limit however, to how much each curve is shifted for a fixed value of $\mu^2$. For example, in figure~\ref{fig:sig_C_cont_varbeta100}, the $\mu^2 = 10^{-38}$ GeV$^2$ case (solid blue) is shifted to higher cross sections when $\beta$ is reduced from $\beta = 0.1509$ to $\beta = 10^{-5}$ (i.e. going from panel 1 to panel 2). But, as the value of $\beta$ is reduced further in the successive panels, the curve does not move. A similar thing happens for the $\mu^2 = 10^{-44}$ GeV$^2$ case (solid red) once $\beta \lesssim 10^{-10}$ (panels 3 and 4). We understand this by noting that once the value of $\beta$ has dropped below the threshold given in~\eqref{eq:rs_freeze}, the behaviour of each curve is given by the Randall-Sundrum result (see~\eqref{eq:sigma_asym}), and is therefore independent of $\beta$.

The increased annihilation cross section of the asymmetric dark matter species in the Gauss-Bonnet braneworld scenario gives rise to an interesting prospect, first pointed out in~\cite{Gelmini:2013awa}: if the cross section is large enough, it is possible that the annihilation rate and in turn the indirect detection signal of asymmetric dark matter could be enhanced with respect to the symmetric signal in the standard scenario, despite the suppressed abundance of the minority dark matter component. This behaviour, which is contrary to the usual expectation, is possible in both the quintessence and scalar-tensor scenarios~\cite{Gelmini:2013awa}, as well as the Randall-Sundrum braneworld model~\cite{Meehan:2014zsa}. Since we have shown that the required annihilation cross section in the GB braneworld model is increased by up to several orders of magnitude, we would expect similar behviour here also.

Using the formalism developed in~\cite{Meehan:2014zsa}, we indicate in figures~\ref{fig:sig_C_cont_varbeta100} and~\ref{fig:sig_C_cont_varbeta10} the regions in the $(\langle\sigma v\rangle,C)$ plane that produce an amplified asymmetric dark matter detection signal (dot-dashed blue curve). To compare our results with experiment, we also show the region excluded by the latest Fermi-LAT data~\cite{Fermi} (dot-dashed purple and magenta curves). Combining the two, the allowed region of parameter space that produces an amplified detection signal is given by
\begin{equation}
\langle\sigma v\rangle^{GR} < \langle\sigma v\rangle\gamma < \langle\sigma v\rangle_{\mathrm{Fermi}},
\end{equation}
where $\langle\sigma v\rangle^{GR}$ is the required annihilation cross section for symmetric dark matter in the standard cosmological scenario (see section~\ref{sec:sdm}) and $\gamma$ is a damping factor that arises from the asymmetry between the particles $\chi$ and antiparticles $\bar{\chi}$, given by (see~\cite{Meehan:2014zsa})
\begin{equation}
\gamma\equiv \frac{2Y_\chi Y_{\bar{\chi}}}{\left(Y_\chi + Y_{\bar{\chi}}\right)^2} = \frac{\omega^2 - C^2}{2\omega^2}.
\end{equation}
Figures~\ref{fig:sig_C_cont_varbeta100} and~\ref{fig:sig_C_cont_varbeta10} show that it is possible to produce an amplified asymmetric detection signal in the Gauss-Bonnet braneworld model, however, the allowed region decreases as the dark matter particle mass drops from $m_\chi = 100$ GeV to $m_\chi = 10$ GeV due to the more stringent Fermi-LAT constraints.

\section{Conclusions}
\label{sec:con}

Relic abundance calculations provide an important test of non-standard cosmological scenarios in the early pre-BBN universe (see~\cite{Gelmini:2009yh} for further discussion). In this article we have revisited the relic abundance investigation in the Gauss-Bonnet braneworld scenario in which a Gauss-Bonnet curvature invariant is added to the Randall-Sundrum braneworld action. A previous investigation by~\cite{PhysRevD.79.103528} found that the dark matter density is suppressed in the GB braneworld model, however, this conclusion is based on a highly contrived assumption that collapses the Randall-Sundrum expansion era, leading to a slower early time expansion law. We find that when this assumption is relaxed, the early time expansion rate can be either faster or slower than the standard expansion law, depending on the model parameters. In turn, the dark matter relic abundance is either enhanced or suppressed by up to several orders of magnitude with respect to the standard cosmology result, respectively. Importantly, when realistic parameter values are chosen, the early time expansion rate is typically faster than the standard expansion law during the era of dark matter decoupling and the resulting relic abundance is enhanced. Moreover, in the limit $\beta\lll 1$ (corresponding to $\mathcal{R}_m \gg 1$) the usual Randall-Sundrum type behaviour is recovered~\cite{PhysRevD.70.083531,PhysRevD.79.115023}.

We have also investigated the GB braneworld effect on asymmetric dark matter species and found that the enhanced annihilation cross section required to provide the observed relic density is capable of producing an amplified annihilation signal with respect to the symmetric signal in the standard cosmological scenario. This effect, which is contrary to the usual expectation, has also been demonstrated in quintessence, scalar-tensor~\cite{Gelmini:2013awa} and Randall-Sundrum braneworld models~\cite{Meehan:2014zsa}.

The implications of the latest Fermi-LAT constraints on the dark matter annihilation cross section have been considered for both the symmetric and asymmetric models. For small $\beta$, corresponding to realistic values for the mass ratio $\mathcal{R}_m$, larger values of $\mu^{2}$ are favoured, suggesting that the Gauss-Bonnet braneworld expansion rate has reduced to the standard expansion law before dark matter decoupling. 

The present investigation is timely because the weak scale cross section relevant to generic relic abundance calculations should be accessible to the next generation of direct and indirect detection experiments~\cite{Bauer:2013ihz}. Therefore, additional constraints and/or an unexpected signal from these experiments could point to new physics in the era prior to BBN. 

Our investigation also has implications for dark matter particle models and scans of supersymmetric parameter space. If the early time expansion rate is in fact slower than the standard scenario, particles which are typically overproduced in the standard cosmology and thus ruled out by relic density constraints, may be rescued in the GB scenario.

\end{document}